# Improvement of Clamond's solution of the Colebrook-White equation: highest accuracy for engineering purposes with one iteration

by

Dr.-Ing. Ernst Große-Dunker[1]

21 March 2025

[1] Unaffiliated pensioner and former Mechanical Engineer, Achim, Germany
E-Mail: ernstgd@gmx.de
ORCID iD: https://orcid.org/0009-0003-4607-702X

## Abstract

The Colebrook-White equation is the widely used basis for the calculation of the friction factor λ for flows in pipes and ducts. Because this equation is implicit in λ, many solutions have been developed to ease the calculation in order to reduce the effort and to reach a sufficient accuracy. Clamond has proposed in 2008 an iterative solution that requires maximally two iterations to obtain the machine double precision. Here an improvement of this solution is presented, that achieves already with one iteration a maximal error of $2.79 \cdot 10^{-7}$, what is more than sufficient for most engineering purposes. This solution is compared in a chart of CPU time versus accuracy with 28 solutions from the literature and in the group of the fastest solutions, that require only two calls of the logarithm function, it proved to be by far the most accurate one.

Keywords: Friction factor, Colebrook-White, Wright omega function, iterative solution

## Introduction

The frictional pressure loss of flows in pipes is of high interest in the design and simulation of chemical and power plants, and of gas, oil and water distribution and transportation systems. The surface of the pipes or ducts may differ from very smooth to very rough, and the flow may differ in a wide range from laminar to turbulent flow. Colebrook [1] has developed an equation for turbulent flow that covers the full range from full smooth to full rough surfaces

$$\frac{1}{\sqrt{\lambda}} = -2\,\lg\left(\frac{k}{C_k} + \frac{C_{Re}}{Re\,\sqrt{\lambda}}\right) \tag{1}$$

where lg denotes the logarithm to the base 10, Re the Reynolds number, and k the relative roughness, i.e. the roughness height divided by the diameter of the pipe or duct, $C_k$ and $C_{Re}$ are constants, determined by Colebrook to $C_k$=3.7 and $C_{Re}$=2.51, and λ is the Weisbach friction factor [2] defined by

$$\Delta p = \lambda \frac{\Delta l}{d} \frac{\rho w^2}{2} \tag{2}$$

where Δp is the pressure drop of the pipe, the ratio Δl/d the geometry factor for long straight pipes with the length Δl and the diameter d, $\frac{1}{2}\rho w^2$ the dynamic pressure of the fluid with the density ρ and the velocity w. Moody [3] has given the validity range for eq. (1) with $4000 \le Re \le 10^8$ and $0 \le k \le 0.05$.

The implicit Colebrook-White equation can be solved by iterative procedures or by approximating explicit solutions. Iterative methods are given in [4–6], comprehensive summaries of explicit approximations may be found in [7–10]. Further on, the Colebrook-White equation can be solved directly with the Lambert W- and the Wright ω-functions, see [11–17], but due to numerical overflows the Lambert W-function has a limited usability on standard computer environments.

Based on the Wright ω-function, Clamond [11] has developed a fast and accurate iterative solution, that receives machine double precision accuracy with maximal two iterations. Here an improvement of this solution is presented, that achieves already with the first iteration an accuracy, that is sufficient for most engineering applications.

All accuracy calculations and CPU time measurements are carried out with a FORTRAN integrated development environment for Windows, the details of the examination of the accuracy and CPU time are given in Appendix A. The solutions from literature used for the comparison with the improved solution are listed in Appendix B, and the FORTRAN source code of the improved solution is given in Appendix C.

## Iterative solution proposed by Clamond 2008 [11]

For solving the Colebrook-White equation (1) with the Wright ω-function it has been rearranged to:

$$\mathrm{R} + \ln(\mathrm{R}) = \mathrm{S} \tag{3}$$

with

$$\mathrm{R} = \mathrm{x} + \mathrm{M} \tag{4}$$

$$\mathrm{S} = \mathrm{N} + \mathrm{M} \tag{5}$$

$$\mathrm{x} = \frac{1}{\sqrt{\lambda}} \frac{\ln(10)}{2} \tag{6}$$

$$\mathrm{M} = \frac{\ln(10)}{2} \frac{\mathrm{k}}{\mathrm{C_k}} \frac{\mathrm{Re}}{\mathrm{C_{Re}}}; \quad \mathrm{C_k} = 3.7; \quad \mathrm{C_{Re}} = 2.51 \tag{7}$$

$$\mathrm{N} = \ln\left(\frac{\ln(10)}{2} \frac{\mathrm{Re}}{\mathrm{C_{Re}}}\right) \tag{8}$$

Eq. (3) can be inverted with the Wright ω-function to

$$\mathrm{R} = \omega(\mathrm{S}) \tag{9}$$

where R is the variable of interest, and after R being determined, x respectively λ can be calculated by means of eq. (4) and (6). For the iterative solution of eq. (9) Clamond applied the Householder method, see [11, 18],

$$\mathrm{x_i} = \mathrm{x_{i-1}} + \mathrm{p}\left[\frac{\left(\frac{1}{\mathrm{f(x)}}\right)^{(\mathrm{p}-1)}}{\left(\frac{1}{\mathrm{f(x)}}\right)^{(\mathrm{p})}}\right]_{\mathrm{x}=\mathrm{x_{i-1}}} \tag{10}$$

where p denotes a whole number greater than 0, and $F^{(p)}$ the $p^{th}$ derivative of the function F with the $0^{th}$ derivative defined as $F^{(0)}=F$. With p=3 and a shift of the ω-function by M the following sequence of equations for the iteration has been received

$$\mathrm{x_i} = \mathrm{x_{i-1}} - \frac{\left(1 + \mathrm{x_{i-1}} + \mathrm{M} + \frac{1}{2}\varepsilon_{\mathrm{i-1}}\right)(\mathrm{x_{i-1}} + \mathrm{M})\,\varepsilon_{\mathrm{i-1}}}{1 + \mathrm{x_{i-1}} + \mathrm{M} + \varepsilon_{\mathrm{i-1}} + \frac{1}{3}\varepsilon_{\mathrm{i-1}}^2} \qquad \text{for i = 1 to n} \tag{11}$$

$$\varepsilon_{\mathrm{i-1}} = \frac{\ln(\mathrm{x_{i-1}} + \mathrm{M}) + \mathrm{x_{i-1}} - \mathrm{N}}{1 + \mathrm{x_{i-1}} + \mathrm{M}} \tag{12}$$

Finally, when the iteration is stopped at i = n

$$\lambda = \left(\frac{1}{\mathrm{x_n}}\right)^2 \left(\frac{\ln(10)}{2}\right)^2 \tag{13}$$

For the initial starting point $x_0$ the equation

$$x_0 = N(Re) + A; \quad A = -0.2 \tag{14}$$

is given. It is stated that the machine double precision is obtained already after one iteration, except for

$$M + N \leq 5700 \tag{15}$$

when the second iteration is required to obtain double precision. This solution requires one one-time logarithm calculation, i.e. ln(Re) for the calculation of N in eq. (8) and $x_0$ in eq. (14), and one repetitive logarithm calculation, i.e. $\ln(x_{i-1}+M)$ for the calculation of $\varepsilon_i$ in eq. (12). Caused by the condition of eq. (15), either one or two iterations are to be processed by a computer program; what means, that for the test grid given in Appendix A, that covers the full validity range of the Colebrook-White equation, an average of 1.63 iterations, respectively of 2.63 logarithm calls, are to be processed.

## Improvement of the first iteration

When running only the first iteration of Clamond's solution, the maximum error is $1.54 \cdot 10^{-4}$, see ID #1 in Table 1. This is already a good accuracy and better than that of several other solutions from literature, see the chapter below. Further on this error can be significantly decreased with a new simple approach for $x_0$, where the initial Reynolds number is a linear function of the Reynolds number $Re_0$ = D (Re + B)

$$x_0 = N(Re_0) = \ln\left(\frac{\ln(10)}{2 \cdot C_{Re}}\, D\,(Re + B)\right) = \ln\left(\frac{\ln(10)}{2 \cdot C_{Re}}\right) + A + \ln(Re + B) \tag{16}$$
$$A = \ln(D)$$

With the numerical values A=–0.2 and B=0 eq. (16) matches Clamond's initial starting point given in eq. (14).

If eq. (16) would be applied as is, it would require, besides the above mentioned calculations of ln(Re) for N in eq. (8) and $\ln(x_{i-1}+M)$ for $\varepsilon_i$ in eq. (12), a third logarithm call, i.e. ln(Re+B) for the calculation of $x_0$. But the call of three logarithms can be avoided in the following way, when considering that, in case of only one iteration, N is used only once for the calculation of $\varepsilon_0$: when calculating in eq. (12) the difference $x_0$-N with eq. (16) and (8), eq. (12) becomes

$$\varepsilon_0 = \frac{\ln(x_0 + M) + A + \ln(1 + B/Re)}{1 + x_0 + M} \tag{17}$$

When taking into account, in advance to the following optimization, the numerical value of B from Table 1, and according to the range of validity the minimal Reynolds number $Re_{min}$ = 4000, then the result is $B/Re < 10^{-6}$, and it can be assumed that $\ln(1+B/Re) \approx B/Re$. Based on this, C is introduced in eq. (17)

$$\varepsilon_0 = \frac{\ln(x_0 + M) + A + C}{1 + x_0 + M} \tag{18}$$

and considered here as a constant close to zero, what avoids the calculation of the third logarithm function. Note, that in the right part of eq. (16) the left and the middle term can be summarized to one constant, as well as the middle and the right term of the numerator in eq.(18).

To improve the accuracy of Clamond's solution with only one iteration, the constants A, B and C are subject of the optimization, the result of which is given in Table 1. The first line, ID #1, represents the first iteration of Clamond's original solution with $x_0$ according to eq. (14). In the first step of the optimization the constant A is optimized and B and C are kept zero. The maximum absolute error $E_{abs}$=4.6·10$^{-7}$ is received with A=-2.0424324, see #2 in Table 1, but the maximum positive and negative errors show, that the error is asymmetrically distributed solely to the positive side, when taking into account, that $E_{neg}$ is at the limit of the FORTRAN double precision and therefore can be considered as zero. This first optimization step reduces the error, that is achieved with Clamond's initial starting point, by the factor 338, see most right column of Table 1. It is noted, that Areerachakul et al. [19] reported about an optimized initial starting point for Clamond's approach, by which they obtained with one iteration an error of 4.545·10$^{-7}$, however, besides this numerical value, no further details about the optimization have been given.

For the further optimization A is kept constant and B and C are adapted to reduce the error. This leads to two almost equivalent results, see #3 and #4 in Table 1, with the effect, that in both cases the error is now evenly distributed between the positive and negative side. It is plausible, that by this shift the maximum absolute error is more or less halved, see the factors of 550 and 552 in the right column for #3 and #4 in comparison to the factor 338 for #2.

**Table 1: Error of optimization for solutions with n=1**
**#1** – Clamond's original solution, **#2** – first optimization step, **#3** and **#4** – further steps of optimization
**A, B, C** – constants, subject of optimization
**$E_{pos}$, $E_{neg}$, $E_{abs}$** – maximum relative positive, negative, and absolute errors

| ID | A | B | C | $E_{pos}$ | $E_{neg}$ | $E_{abs}$ | $E_{abs}$(#1) / $E_{abs}$ |
|---|---|---|---|---|---|---|---|
| **#1** | -0.2 | 0.0 | 0.0 | 1.54·10$^{-4}$ | -1.57·10$^{-15}$ | 1.54·10$^{-4}$ | 1 |
| **#2** | -2.0424324 | 0.0 | 0.0 | 4.56·10$^{-7}$ | -1.17·10$^{-15}$ | 4.56·10$^{-7}$ | 338 |
| **#3** | -2.0424324 | 0.0033774 | 0.0 | 2.80·10$^{-7}$ | -2.80·10$^{-7}$ | 2.80·10$^{-7}$ | 550 |
| **#4** | -2.0424324 | 0.0 | -6.0·10$^{-7}$ | 2.79·10$^{-7}$ | -2.79·10$^{-7}$ | 2.79·10$^{-7}$ | 552 |

The three solutions, #1, #2 and #4, have exactly the identical number of operations, only #3 requires one addition more, that is the addition Re+B in eq. (16). The advantage of improvement #4 over #3 is, that it is a little bit more accurate and that it requires one addition less. Therefore, the solution #4 is considered to provide the best improvement with regard to accuracy and CPU time and it is used for the following comparison and labelled with Clamond 1it opt.

## Comparison of improved solution with solutions from literature

The above found improvement of the first iteration of Clamond's solution, #3 in Table 1, is compared with 28 solutions from 13 sources in the literature. Included herein are three variants of Clamond's solution with the original initial starting point of eq. (14): • with a fixed number of one iteration, • with a fixed number of two iterations, • with the condition given in eq. (15), that leads to an average of 1.63 iterations for the test grid given in Appendix A. The details of all solutions and their labels used hereunder are given in Appendix B.

### Verification of the calculated accuracies

In order to verify the programmed solutions from the literature with regard to the accuracy, the calculated errors of these solutions are compared with the reported errors in the original references. In Table 3 in Appendix B the error $E_{abs}$, the validity range of $E_{abs}$ and the constant $C_k$ from the 13 references are given, the column 'Deviating range' contains the summarized information about the deviation of the range of $E_{abs}$ validity from the validity range of the Colebrook-White equation.

The comparison of calculated and reported errors in Table 2 shows, that the calculated errors are in good agreement with the reported ones, minor differences are in several cases explainable by the use of deviating validity ranges in this comparison and in the references. Besides this, the following three deviations in Table 2 are pointed out: • For the solution Praks 20b-30 the error of $2.4 \cdot 10^{-7}$ is reported in [15], but cannot be reproduced with the given equations, instead, the calculated error is with $3.9 \cdot 10^{-6}$ about 16 times larger than the reported one. On the other hand, the error for Praks 20b-29 from the same reference can be exactly reproduced. • For all three solutions Praks 18-28.1, Praks 18-28.2 and Praks 18-28.3 in [4] an error of $6.17 \cdot 10^{-4}$ is reported for the full validity range of the Colebrook-White equation, but the calculated errors are with $1.5 \cdot 10^{-3}$ to $1.7 \cdot 10^{-3}$ about 2.6 times larger. However, looking at figures 6, 7 and 8 of the reference, the lower limits of the range of validity are shown with $10^4$ on the Re-axis and $10^{-6}$ on the k-axis, and when applying test wise these limits, the calculated errors match with about $6.16 \cdot 10^{-4}$ the reported one. • Clamond reports in [11] a double precision accuracy for his solution, but the calculated accuracy for Clamond orig is 'only' 12.25 MAD. From a practical point of view, this deviation is of no interest, because an accuracy of more than 12 digits is far more than required for any application.

**Relative CPU time (RCT) versus minimal accurate digits (MAD)**

The results of this comparison are given in Table 2 as numerical data and in Figure 1 as a graph of the RCT (Relative CPU Time) versus MAD (Minimal Accurate Digits), both in detail explained in Appendix A. The RCT is the measured CPU time of the solution in relation to the measured CPU time of the reference solution, which is here Clamond 1it opt. The MAD is a meaningful linear and absolute scale, that is not related to a reference solution and that indicates the number of minimal accurate digits for the calculated friction factor of the solution.

The RCTs given in Table 2 and in Figure 1 are the averages of 6 runs. When comparing for each solution the 6 individual RCTs with its average of the 6 runs, they all are within ±1% of the average, what indicates a very high reproducibility of the CPU time measurement. The solution Clamond 1it opt is taken as the reference for the RCT, therefore its RCT is per definition 100%. For a better resolution of the results in the interesting RCT range, the RCT scale in Figure 1 is limited between 80% and 180% RCT, and therefore the solutions Vatankhah orig, Sonnad CFA orig and Chen with more than 200% RCT, are not shown in Figure 1.

From Table 2 it can be seen, that the fastest first 15 solutions all require two calls of the logarithm functions, and the differences in the RCT are caused by the different processing efforts for the basic arithmetic operations. Consequently, their RCT values are close together in the range of 97% to 110%, but their errors vary widely between $1.0 \cdot 10^{-2}$ and $2.8 \cdot 10^{-7}$, respectively the accuracies vary between 2.00 to 6.55 MAD. All solutions further down in Table 2 require at least either 1 logarithm plus 1 power-function or 3 logarithm functions, they are in the range of 141% to 258% RCT, only 4 solutions of these provide a better accuracy than Clamond 1it opt, but require at least 145% RCT.

The solutions Vatankhah orig and Sonnad CFA orig with the original equations have a RCT of 240% and 258%, while the solutions Vatankhah and Sonnad CFA with rearranged equations according to Appendix B have a RCT of 108% and 146%, i.e. this rearrangement, that avoids in case of Vatankhah 1 power- and 1 ln-function and in case of Sonnad CFA 1 power-function, reduces the RCT for Vatankhah by about 130% and for Sonnad CFA by about 100% without a loss of accuracy.

The comparison of the accuracy between Vatankhah and Vatankhah orig shows, that Vatankhah is more accurate, i.e. the calculated error is by the factor 2.8 lower than that of Vatankhah orig, what is solely caused by the rounding of the constants to 4 respectively 5 decimal places in the reference, see [16] or eq. (47) in Appendix B, while in the solution Vatankhah the mathematical equations for the constants are retained, see eq. (48) in Appendix B, and computed as named constants in the FORTRAN FUNCTION.

In order to identify the solutions, that provide the best relation of accuracy to CPU time the waterfall method is applied to column MAD in Table 2, sorted by RCT, i.e. symbolically the water is falling down in column MAD from the first line to the last, and all rows are getting wet and build a new waterfall edge whose MAD is larger than that of the previous edge. The rows with a MAD less than that of the previous edge are covered by it and don't get wet, meaning that these solutions provide no advantage, because they have a higher RCT but a lower MAD. In total there are the following 5 waterfall edges marked in bold in Table 2 and Figure 1: Praks 20a-3, Praks 20b-29, Clamond 1it opt, Clamond orig and Clamond 2it. But only the first three are of interest, of which Clamond 1it opt provides with 6.55 MAD highest accuracy. The last two provide an unnecessary high level of accuracy and require at least 45% more RCT than Clamond 1it opt.

But also, with a simpler view, when considering all solutions with two ln-functions to be equivalent with regard to the CPU time, Clamond 1it opt provides by far the best accuracy of those.

**Table 2: Comparison of improvement with solutions from literature**

- Sorted by RCT • MAD waterfall edges marked in bold
- ([1]) = Average value for test grid
- From Colebrook-White deviating range of validity: (–) = going below, (+) = exceeding, details see Table 3

| **Label of solution** (see Appendix B) | $E_{abs}$ (calculated here) | **MAD** | **RCT** | $E_{abs}$ (from references) | **Number of functions** | |
|---|---|---|---|---|---|---|
| | | | | | ln(x) | $x^y$ |
| **Praks 20a-3** | $3.7\cdot10^{-4}$ | **3.44** | 97.4% | $3.66\cdot10^{-4}$ | 2 | 0 |
| **Praks 20b-29** | $1.2\cdot10^{-5}$ | **4.92** | 97.9% | $1.20\cdot10^{-5}$ | 2 | 0 |
| Praks 20a-4 | $8.2\cdot10^{-5}$ | 4.09 | 98.1% | $8.08\cdot10^{-5}$ | 2 | 0 |
| Praks 20a-2 | $1.0\cdot10^{-3}$ | 3.00 | 98.3% | $1.01\cdot10^{-3}$ | 2 | 0 |
| Biberg-14 | $1.5\cdot10^{-3}$ | 2.82 | 98.8% | $1.53\cdot10^{-3}$ | 2 | 0 |
| Clamond 1it | $1.5\cdot10^{-4}$ | 3.81 | 99.9% | n. a. | 2 | 0 |
| **Clamond 1it opt** | $2.8\cdot10^{-7}$ | **6.55** | 100% | n. a. | 2 | 0 |
| Sonnad – | $1.0\cdot10^{-2}$ | 2.00 | 100% | $1\cdot10^{-2}$ (–) | 2 | 0 |
| Lamri-14 | $1.5\cdot10^{-3}$ | 2.82 | 101% | $1.49\cdot10^{-3}$ (+) | 2 | 0 |
| Biberg-15 | $6.1\cdot10^{-5}$ | 4.22 | 101% | $6.10\cdot10^{-5}$ | 2 | 0 |
| Lamri-15 | $3.8\cdot10^{-4}$ | 3.42 | 102% | $4.00\cdot10^{-4}$ (+) | 2 | 0 |
| Praks 20b-30 | $3.9\cdot10^{-6}$ | 5.40 | 102% | $2.40\cdot10^{-7}$ | 2 | 0 |
| Lamri-16 | $1.9\cdot10^{-5}$ | 4.72 | 104% | $2.00\cdot10^{-5}$ (+) | 2 | 0 |
| Vatankhah | $9.9\cdot10^{-6}$ | 5.01 | 108% | $2.80\cdot10^{-5}$ (+) | 2 | 0 |
| Sergh-3 | $4.5\cdot10^{-3}$ | 2.35 | 110% | $1.98\cdot10^{-3}$ (–,+) | 2 | 0 |
| Haaland | $1.4\cdot10^{-2}$ | 1.85 | 141% | $1.5\cdot10^{-2}$ | 1 | 1 |
| Swamee | $3.4\cdot10^{-2}$ | 1.47 | 143% | $1\cdot10^{-2}$ (–) | 1 | 1 |
| Sonnad LA | $3.6\cdot10^{-6}$ | 5.44 | 145% | $3.64\cdot10^{-6}$ (–) | 3 | 0 |
| **Clamond orig** | $5.6\cdot10^{-13}$ | **12.25** | 145% | $1\cdot10^{-15}$ (+) | 2.63 ([1]) | 0 |
| Sonnad CFA | $1.0\cdot10^{-12}$ | 11.98 | 146% | $1.04\cdot10^{-12}$ (–) | 3 | 0 |
| Sergh-2 | $3.1\cdot10^{-5}$ | 4.50 | 149% | $2.3\cdot10^{-5}$ (–,+) | 3 | 0 |
| Zigrang | $1.1\cdot10^{-3}$ | 2.94 | 157% | $1.10\cdot10^{-3}$ (–) | 3 | 0 |
| Praks 18-28.2 | $1.7\cdot10^{-3}$ | 2.77 | 165% | $6.17\cdot10^{-4}$ | 3 | 0 |
| Praks 18-28.1 | $1.7\cdot10^{-3}$ | 2.76 | 166% | $6.17\cdot10^{-4}$ | 3 | 0 |
| **Clamond 2it** | $1.1\cdot10^{-15}$ | **14.94** | 167% | $1\cdot10^{-15}$ (+) | 3 | 0 |
| Praks 18-28.3 | $1.5\cdot10^{-3}$ | 2.82 | 173% | $6.17\cdot10^{-4}$ | 3 | 0 |
| Vatankhah orig | $2.7\cdot10^{-5}$ | 4.56 | 240% | $2.80\cdot10^{-5}$ (+) | 3 | 1 |
| Sonnad CFA orig | $1.0\cdot10^{-12}$ | 11.98 | 252% | $1.04\cdot10^{-12}$ (–) | 3 | 1 |
| Chen | $4.63\cdot10^{-3}$ | 2.49 | 258% | $4.65\cdot10^{-3}$ (–,+) | 2 | 2 |

The solution Clamond 1it, i.e. Clamond's original approach with only one iteration, is more accurate than 14 solutions of this comparison, its comparison with Clamond 1it opt shows, that both require the same RCT of 100%, but Clamond 1it opt is about 2.7 MAD more accurate.

**Figure 1: Comparison of solutions from literature with improved solution – RCT versus MAD**

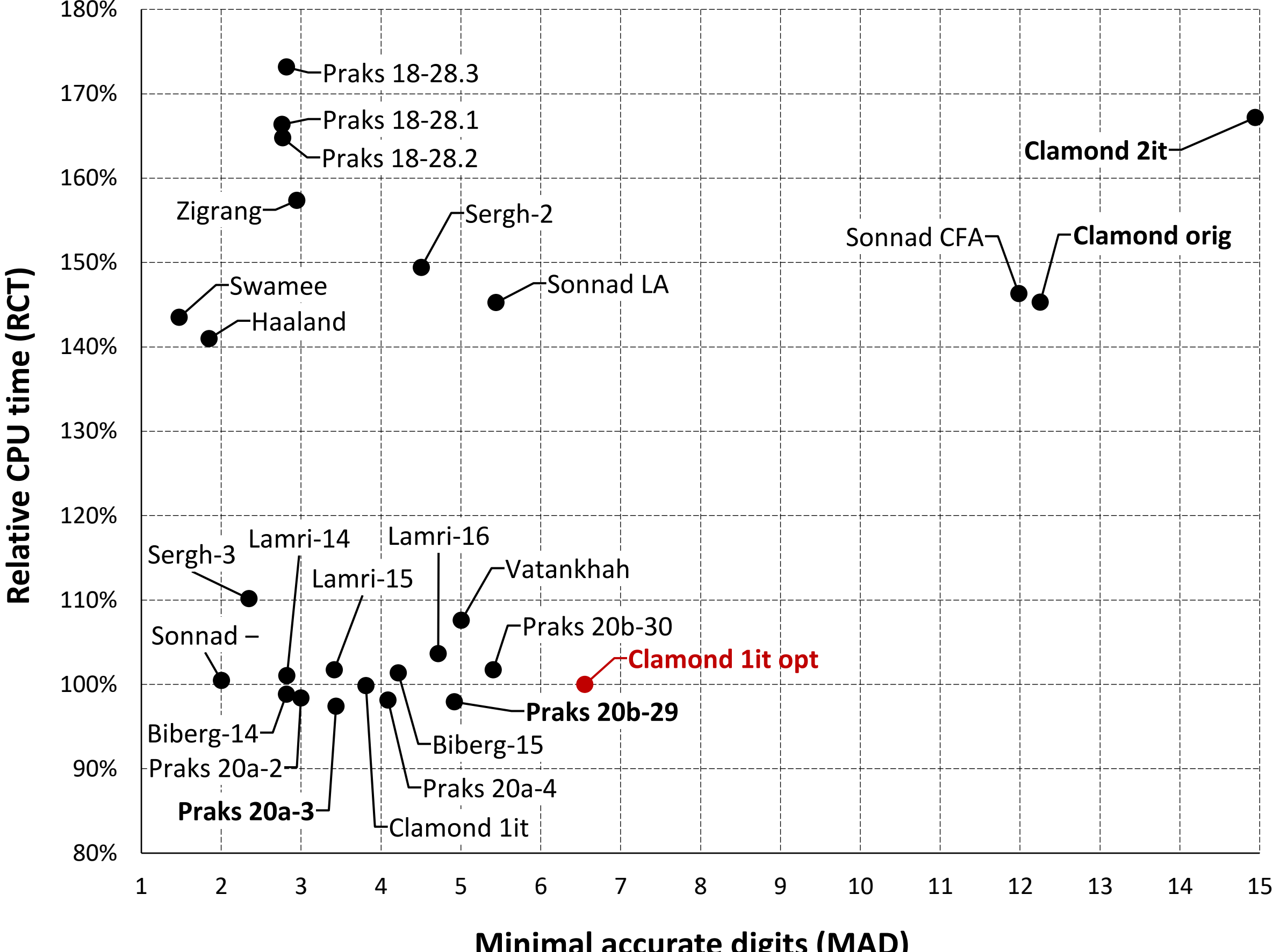

Figure 1 shows the RCT (relative CPU time) of the compared solutions versus the MAD (minimal accurate digits). In the range of about 100% RCT there are 15 solutions that require the calculation of two logarithms, the differences in RCT are caused by the different processing efforts of the basic arithmetic operations. In this group Clamond 1it opt is the one with an outstanding MAD of 6.55, what is sufficient for most engineering purposes. All other solutions require more than 140% RCT, eight of them with a MAD below that of Clamond 1it opt, three of them with a MAD above 12, but this high level of accuracy is not required in engineering applications.

## Conclusion

Clamond's iterative solution of the Colebrook-White equation has been improved by an optimization of the initial starting point. This improvement provides already with only one iteration a minimal accuracy of 6.55 digits, respectively a maximal error of $2.79 \cdot 10^{-7}$, most likely sufficient for all engineering purposes. In a comparison with 28 solutions from the literature it demonstrates its competitiveness: it belongs to the group of the fastest solutions, that require only 2 logarithm functions, and it is by far the most accurate one of this group.

# Appendix A: Verification of the accuracy and CPU load

## Test grid

For the verification of the accuracy of the solutions a test grid is applied that covers the full validity range of Colebrook-White equation ($0 \le k \le 0.05$, $4000 \le Re \le 10^8$). In the k versus log(Re) plane of the test grid the k-axis and log(Re)-axis are each divided into 1000 regular intervals. Including the upper and lower limits, the axes consist of 1001 points each and the test grid of n=1,002,001 points. Even if in the literature authors have noted a different range of validity, the full test grid is applied to all compared solutions. A mismatch between the accuracy given in the literature references and the one calculated here may be explainable by this deviation, or by a test grid with significant fewer points.

## Error criteria

The relative error of a single point in the test grid is

$$E_i = (\lambda_i - \lambda_{i,ref})/\lambda_{i,ref} \tag{19}$$

with $\lambda_i$ – value of the equation under examination and $\lambda_{i,ref}$ – value of the reference equation, which is the solution Clamond 4it by Clamond [11] with a fixed number of 4 iterations, see Appendix B, it is used only as the reference for the accuracy verification. In [11] the roughness factor $C_K$=3.7 is used. If the solution to be verified uses a different value instead, see most right column in Table 3, the constant is adapted in this reference. Clamond 4it is not considered for the RCT versus MAD comparison, see below.

The maximum relative positive, maximum relative negative, and maximum absolute relative errors are defined as

$$E_{pos} = MAX(0, E_1, \dots, E_i, \dots, E_n) \tag{20}$$

$$E_{neg} = MIN(0, E_1, \dots, E_i, \dots, E_n) \tag{21}$$

$$E_{abs} = MAX(E_{pos}, |E_{neg}|) \tag{22}$$

The only quality criteria applied here for the assessment of the accuracy of the solutions under examination is the maximum absolute relative error $E_{abs}$, because 1) the difference between two mathematical functions is verified and the error is therefore a functional one and not caused by noise or scatter, and 2) the error of the examined solutions is very small.

## Relative CPU time (RCT) versus minimal accurate digits (MAD)

The comparison of the performance of the examined solutions of the Colebrook-White equation is presented in a chart of the relative CPU time (RCT) versus minimal accurate digits (MAD). The relative CPU time (RCT) is defined as

$$RCT = \frac{T}{T_{ref}} \tag{23}$$

where T is the CPU time that the examined equation takes and $T_{ref}$ is the CPU time that the reference equation Clamond 1it opt takes, see Appendix B. In order to avoid the effect of outliers in the CPU time measurement, the sum of n time measurements is taken for the calculation of T and $T_{ref}$

$$T = \sum_{i=1}^{n} t_i \quad ; \quad T_{ref} = \sum_{i=1}^{n} t_{ref,i} \tag{24}$$

where $t_i$ denotes the CPU time of a single run. The minimal accurate digits (MAD) are defined as

$$\mathrm{MAD} = \lg\left(\frac{1}{\mathrm{E_{abs}}}\right) \tag{25}$$

where lg denotes the logarithm to the base 10, MAD represents the minimal number of accurate digits for a given maximum absolute relative error $E_{abs}$. For example, if $E_{abs}=1.0\cdot10^{-3}$ then the MAD is 3.0, meaning that at minimum the first 3.0 digits are accurate. Or, in a more general term, the last digit of the whole number of $10^{MAD}$ is still correct.

For the verification of the accuracy and the CPU time the free version of the Silverfrost IDE Plato (version 5,6,0,0) with the FORTRAN 95 Personal Edition (version 8.95) in Win32 environment is used. The CPU time is determined with the Plato option 'Plant Timing Information'. With this option the times, the CPU takes to execute the different FORTRAN FUNCTIONs, are recorded in a file; the transfer time from the calling to the called program is assigned to the calling program, the return transfer time is assigned to the called program, i.e. the time T of the examined program includes the time to execute the program and the transfer time back to the calling program. For all calculations the Plato option 'Set Default Real kind = 2' has been set, i.e. Double Precision with an accuracy of 15 digits is the default for all floating-point variables. No Plato option for compiler optimization has been set.

The RCT is determined in the following way: An executable with all examined solutions is build, that runs the test grid for 30 times to get reasonable large numerical values for the CPU times. This executable is run 6 times in order to eliminate noise or outliers in the CPU time measurement. Then the RCT is calculated with the recorded data according to eq. (23) and (24) for each FUNCTION.

The reproducibility of the CPU time measurement is very high, i.e. when calculating for each of the 6 runs the RCT for the FUNCTIONs and comparing them with the average RCT according to eq. (23), all deviations are within ±1% RCT.

**Applied rules for programming the solutions**

The following rules for the programming of the FORTRAN FUNCTIONs are applied to reduce the CPU time:

- Calculated named constants are used as much as possible, because their numerical values are calculated during compilation, not at run time.
- Divisions by constants are changed into multiplication with constants of the inverse.
- Power to a whole number is not applied. First, if applicable, the Horner scheme is applied, second the remaining power-functions are resolved by multiplications.
- DO loops are not applied.
- If expressions are repeatedly used, they are only calculated once and saved as internal variables for reuse.

# Appendix B: Solutions from literature for comparison

In this appendix the solutions from literature used for the comparison with the developed solution **Clamond 1it opt** are listed. For a better identification and differentiation each solution has been given an underlined label as the reference for the comparison. For these solutions the ranges of validity, the errors $E_{abs}$ and the constant $C_k$ to the relative roughness are listed in Table 3 at the end of this appendix.

### Swamee and Jain (1976) / [20] eq. 20

Swamee and Jain applied Colebrook's method with the explicit equation for the friction factor of smooth pipes given by Colebrook [1]. With the factor −2.0 in front of the logarithm function they received

$$\frac{1}{\sqrt{\lambda}} = -2.0\,\lg\left(\frac{5.74}{Re^{0.9}} + \frac{k}{3.7}\right) \tag{26}$$

**Swamee** (26)

### Chen (1979) / [21] eq. 7

Chen has referenced the Colebrook-White equation with $C_k$=3.7065 and $C_{Re}$=2.5226, and proposed a one-step iterative solution with an optimized initial guess.

$$\frac{1}{\sqrt{\lambda}} = -2.0\,\lg\left[\frac{k}{3.7065} - \frac{5.0452}{Re}\lg\left(\frac{k^{1.1098}}{2.8257} + \frac{5.8506}{Re^{0.8981}}\right)\right] \tag{27}$$

**Chen** (27)

### Zigrang and Sylvester (1982) / [22] eq. 12

Zigrang and Sylvester proposed a two-step iterative solution with an optimized initial guess.

$$\frac{1}{\sqrt{\lambda}} = -2.0\,\lg\left\{\frac{k}{3.7} - \frac{5.02}{Re}\lg\left[\frac{k}{3.7} - \frac{5.02}{Re}\lg\left(\frac{k}{3.7} + \frac{13}{Re}\right)\right]\right\} \tag{28}$$

**Zigrang** (28)

### Haaland (1983) / [23] eq. 5

Haaland applied Colebrook's method with the explicit equation for the friction factor for smooth pipes given by Colebrook in [1]. With the factor −1.8 in front of the logarithm function he received

$$\frac{1}{\sqrt{\lambda}} = -1.8\,\lg\left[\left(\frac{k}{3.7}\right)^{1.11} + \frac{6.9}{Re}\right] \tag{29}$$

**Haaland** (29)

### Serghides (1984) / [24] eq. 2 and 3

Serghides applied Steffensen's accelerated convergence technique to an iterative solution of the Colebrook-White equation and proposed two solutions.

$$\lambda = \left(A - \frac{(B-A)^2}{C - 2B + A}\right)^{-2} \tag{30}$$

**Sergh-2** (30)

$$\lambda = \left(4.781 - \frac{(A-4.781)^2}{B - 2A + 4.781}\right)^{-2} \tag{31}$$

**Sergh-3** (31)

$$A = -2\,\lg\left(\frac{k}{3.7} + \frac{12}{Re}\right);\ B = -2\,\lg\left(\frac{k}{3.7} + \frac{2.51}{Re}A\right);\ C = -2\,\lg\left(\frac{k}{3.7} + \frac{2.51}{Re}B\right) \tag{32}$$

### Sonnad and Goudar (2007) / [25] eq. 12 and 13

Sonnad and Goudar have developed a mathematically equivalent representation of the Colebrook-White equation to compute the friction factor.

$$\frac{1}{a\sqrt{\lambda}} = \ln\left(\frac{d}{q}\right) + \delta$$
$$s = bd + \ln(d);\ q = s^{\,s/(s+1)};\ g = bd + \ln\left(\frac{d}{q}\right);\ z = \ln\left(\frac{q}{g}\right) \tag{33}$$

$$a = \frac{2}{\ln(10)};\ b = \frac{k}{3.7};\ d = \frac{\ln(10)}{2}\frac{Re}{2.51} \tag{34}$$

with δ according to eq. (36), (37) and (38). For eq. (33) three logarithm and one power-function are to be calculated. But, when applying the logarithm rules ln(x/y) = ln(x) – ln(y) and $\ln(x^y)$ = y ln(x) to the logarithm and the power-functions in the eq. (33), the power-function can be removed, while the number of logarithm functions stays three.

$$\frac{1}{a\sqrt{\lambda}} = \ln(d) - Q + \delta$$
$$s = bd + \ln(d);\ Q = \frac{s}{s+1}\ln(s);\ g = bd + \ln(d) - Q;\ z = Q - \ln(g) \tag{35}$$

In order to evaluate the CPU time savings made by the rearrangement in eq. (35) also the original equation for $\delta_{CFA}$, which is the most accurate one of the three alternatives, is programmed with three logarithm and one power-function. All in all, four solutions of this approximation are considered for the comparison, eq. (35) with eq. (36), (37) and (38), respectively, and eq. (33) with eq. (38).

| | | |
|---|---|---|
| $\delta = 0$ | **Sonnad –** | (36) |
| $\delta = \delta_{LA} = \frac{g}{g+1} z$ | **Sonnad LA** | (37) |
| $\delta = \delta_{CFA} = \delta_{LA}\left[1 + \frac{z/2}{(g+1)^2 + (2g-1)z/3}\right]$ | **Sonnad CFA** | (38) |
| eq. (33) with eq. (38) | **Sonnad CFA orig** | (39) |

**Clamond (2008) / [11] eq. 17**

For the comparison three solutions based on Clamond's work are chosen: besides the original solution with the condition given by eq. (15) for the second iteration, two solutions with a fixed number of one and two iterations, respectively. The solution Clamond 4it with a fixed number of 4 iterations, eq. (43), is only used as the reference equation for the accuracy, but not for the comparison.

| | | |
|---|---|---|
| eq. (10) to (15) | **Clamond orig** | (40) |
| eq. (10) to (13) for $n = 1$ | **Clamond 1it** | (41) |
| eq. (10) to (13) for $n = 2$ | **Clamond 2it** | (42) |
| eq. (10) to (13) for $n = 4$ | **Clamond 4it** | (43) |

**Biberg (2017) / [12] eq. 14 and 15**

The exact solution of the Colebrook-White equation with the Lambert W-function is approximated by Biberg with a truncated series expansion. Two solutions, eq. 14 with the first three terms, and eq. 15 with the first five terms of the series, have been proposed by Biberg.

$$\lambda = \left\{ a \left[ \ln\left(\frac{Re}{a\,b}\right) + G(x) \right] \right\}^{-2}$$
$$a = 2/\ln(10)\ ;\ b = 2.51\ ;\ c = 3.7$$
$$x = \ln\left(\frac{Re}{a\,b}\right) + \frac{Re\,k}{a\,b\,c};\ z = \frac{1}{x} \tag{44}$$

$$G(x) = \ln(x)\,(z-1) \tag{45}$$

**Biberg-14** (45)

$$G(x) = \ln(x)\left[\big((z-1)z+1\big)z - 1 + \frac{z^2}{6}\ln(x)\big(3 + z(2\ln(x) - 9)\big)\right] \tag{46}$$

**Biberg-15** (46)

**Vatankhah (2018) / [16] eq. 32**

Vatankhah has developed five approximate analytical solutions for the Colebrook-White equation that require one power-function and two or three logarithms, respectively. For this comparison the most accurate solution, i.e. eq. 32 in [16], with three logarithms has been considered.

$$\frac{1}{0.8686\,\sqrt{\lambda}} = \ln\left[\frac{0.3984\,Re}{(0.8686\,s)^{\frac{s}{s+r}}}\right]$$
$$s = 0.12363\,Re\,k + \ln(0.3984\,Re)$$
$$r = 1 + \cfrac{1}{\cfrac{2\,(1+s)}{\ln(0.8686\,s)} - \cfrac{1+4\,s}{3\,(1+s)}} \tag{47}$$

**Vatankhah orig** (47)

By applying the logarithm rules to eq. (47), the power-function can be removed and the number of logarithm functions can be reduced to two, and it becomes

$$\frac{1}{A\,\sqrt{\lambda}} = \ln\left(\frac{Re}{C_{Re}}\right) - \frac{s}{s+r}\ln(A\,s)$$
$$s = \frac{Re\,k}{A\,C_{Re}\,C_K} + \ln\left(\frac{Re}{C_{Re}}\right)$$
$$r = 1 + \cfrac{1}{\cfrac{2\,(1+s)}{\ln(A\,s)} - \cfrac{1+4\,s}{3\,(1+s)}}$$
$$A = \frac{2}{\ln(10)};\ C_K = 3.71;\ C_{Re} = 2.51 \tag{48}$$

**Vatankhah** (48)

In eq. (48) the expressions for the constants are retained to avoid round off effects. In order to evaluate the CPU time savings made by the rearrangement of eq. (48), both solutions, Vatankhah orig and Vatankhah, are considered for this comparison.

**Praks and Brkić (2018) / [4] eq. 28 with part 1, 2, 3 of eq. 27**

Praks and Brkić proposed iterative procedures that only need two steps of iteration, which are:

$$\frac{1}{\sqrt{\lambda}} = -2\,\lg\left(\frac{k}{3.7} + \frac{2.51}{Re}\frac{1}{\sqrt{\lambda_1}}\right)$$
$$\frac{1}{\sqrt{\lambda_1}} = -2\,\lg\left(\frac{k}{3.7} + \frac{2.51}{Re}\frac{1}{\sqrt{\lambda_0}}\right) \tag{49}$$

where $\lambda$ is the final value after the second iteration. In order to receive a sufficient accuracy with only two iterations, the emphasis is put onto the initial starting point $\lambda_0$. The following three alternatives for $\lambda_0$ are proposed:

|  |  |  |
|---|---|---|
| $\frac{1}{\sqrt{\lambda_0}} = 8 - \frac{2\,A}{2 - A\,B}$ | **Praks 18–28.1** | (50) |
| $\frac{1}{\sqrt{\lambda_0}} = 8 - A - \frac{A^2 B}{2}$ | **Praks 18–28.2** | (51) |
| $\frac{1}{\sqrt{\lambda_0}} = 8 - \frac{6\,A - 3\,A^2 B}{6 - 6\,A\,B + A^2\,C}$ | **Praks 18–28.3** | (52) |
| $A = 8 + 2\,\lg\left(\frac{k}{3.7} + \frac{16}{Re}\right);\ B = \frac{-74914381.46}{\nabla^2}$<br>$C = \frac{1391459721232.67}{\nabla^3};\ \nabla = 74205.5 + 1000\,k\,Re$ | | (53) |

**Praks and Brkić (2020) / [26] eq. 2, 3 and 4**

In the initial study [17] the authors developed several approximations based on the Wright ω-function. After the discussion by Niazkar three equations with optimized constants were given in [26]:

|  |  |  |
|---|---|---|
| $\lambda = \left(\frac{1}{0.86902384\,(B + Y)}\right)^2$<br>$A = \frac{Re\,k}{8.11718121};\ B = \ln(Re) - 0.7829415$<br>$x = A + B;\ Y = \ln(x)\left(\frac{1}{x} - 1\right)$ | **Praks 20a–2** | (54) |
| $\lambda = \left(\frac{1}{0.868585\,(B + Y)}\right)^2$<br>$A = \frac{Re\,k}{8.099752};\ B = \ln(Re) - 0.78157$<br>$x = A + B;\ Y = \ln(x)\left(\frac{1.04796}{x + 0.36322} - 1\right)$ | **Praks 20a–3** | (55) |
| $\lambda = \left(\frac{1}{0.868558\,(B + Y)}\right)^2$<br>$A = \frac{Re\,k}{8.0861744};\ B = \ln(Re) - 0.77898$<br>$x = A + B;\ C = \ln(x);\ Y = \frac{1.011746\,C}{x} + \frac{C - 2.3872}{x^2} - C$ | **Praks 20a–4** | (56) |

**Praks and Brkić (2020) / [15] eq. 29 and 30**

The authors present several solutions based on the asymptotic series expansion of the Wright ω-function and symbolic regression. From this study the fastest solution, eq. 29, and the most accurate one, eq. 30, have been selected for the comparison.

|  |  |  |
|---|---|---|
| $\lambda = \left(\frac{1}{0.8685972\ (B + Y)}\right)^2$<br>$A = \frac{Re\ k}{8.0897};\ B = \ln(Re) - 0.779626$<br>$x = A + B;\ C = \ln(x);\ Y = \frac{C}{x - 0.5588\ C + 1.2079} - C$ | **<u>Praks 20b–29</u>** | (57) |
| $\lambda = \left(\frac{1}{0.868589\ (B + Y - \xi)}\right)^2$<br>$A = \frac{Re\ k}{8.088387};\ B = \ln(Re) - 0.7793975$<br>$x = A + B;\ C = \ln(x);\ Y = \frac{C}{x - 0.5564\ C + 1.207} - C$<br>$\xi = \frac{x\ Y^2 + 3.0636\ x\ Y + 18.58}{19.5\ (Y^2\ x^2 + x^3) + 169.9\ Y^2 + 1260\ x + 18178}$ | **<u>Praks 20b–30</u>** | (58) |

**Lamri and Easa (2022) / [27] eq. 14, 15 and 16**

Lamri and Easa have applied the Lagrange inversion theorem to the Colebrook-White equation and receive a series expansion. Three solutions with 2, 3 and 4 series terms have been analysed and optimized, all require only two logarithm calls, i.e. ln(Re) and ln(d).

|  |  |  |
|---|---|---|
| $\frac{1}{\sqrt{\lambda}} = b + 2\ \lg(d) \left(-1 + \frac{0.8645}{d}\right)$ | **<u>Lamri–14</u>** | (59) |
| $\frac{1}{\sqrt{\lambda}} = b + 2\ \lg(d) \left\{-1 + \frac{0.862}{d}\left[1 + \frac{1}{d}\lg\left(\frac{d}{e^2}\right)\right]\right\}$ | **<u>Lamri–15</u>** | (60) |
| $\frac{1}{\sqrt{\lambda}} = b + 2\ \lg(d) \left\{-1 + \frac{0.8682}{d}\left[1 + \frac{1}{d}\lg\left(\frac{d}{e^2}\right)\right] + \frac{0.161}{d^3}[1 + C_1\ \ln(d)] \cdot [4 + C_2\ \ln(d)]\right\}$ | **<u>Lamri–16</u>** | (61) |
| $b = 2\ \lg\left(\frac{Re}{2.51}\right);\ d = \frac{Re\ k}{2.51 \cdot 3.7};\ C_1 = \frac{4}{\sqrt{33} - 9};\ C_2 = \frac{\sqrt{33} - 9}{3}$ | | (62) |

**Table 3: Solutions from literature: Label, range of validity, reported maximum error $E_{abs}$, and $C_k$**

From [1, 3] deviating range of validity: (–) = going below, (+) = exceeding

(*) Accuracy is Double Precision: for the FORTRAN compiler Double Precision is given with 15 digits

(#) $C_k$=3.7065 and $C_{Re}$= 2.5226 are referenced in [21]

| Literature | Label | Deviating range | from references | | | | | |
|---|---|---|---|---|---|---|---|---|
| | | | **Range of validity for $E_{abs}$** | | | | **$E_{abs}$** | **$C_k$** |
| | | | $k_{min}$ | $k_{max}$ | $Re_{min}$ | $Re_{max}$ | | |
| Colebrook [1, 3] | none | | 0 | 0.05 | 4000 | $10^8$ | 0.0 | 3.7 |
| [20] | Swamee | (–) | $10^{-6}$ | 0.01 | 5000 | $10^8$ | $1.0 \cdot 10^{-2}$ | 3.7 |
| [21] | Chen | (–,+) | $5 \cdot 10^{-7}$ | 0.05 | 4000 | $4 \cdot 10^8$ | $4.651 \cdot 10^{-3}$ | 3.7065 (#) |
| [22] | Zigrang | (–) | $4 \cdot 10^{-5}$ | 0.05 | 4000 | $10^8$ | $1.1 \cdot 10^{-3}$ | 3.7 |
| [23] | Haaland | | 0 | 0.05 | 4000 | $10^8$ | $1.5 \cdot 10^{-2}$ | 3.7 |
| [24] | Sergh-2 | (–,+) | $4 \cdot 10^{-5}$ | 0.05 | 2500 | $10^8$ | $2.3 \cdot 10^{-5}$ | 3.7 |
| | Sergh-3 | | | | | | $1.98 \cdot 10^{-3}$ | |
| [25] | Sonnad – | (–) | $10^{-6}$ | 0.05 | 4000 | $10^8$ | $1.00 \cdot 10^{-2}$ | 3.7 |
| | Sonnad LA | | | | | | $3.64 \cdot 10^{-6}$ | |
| | Sonnad CFA | | | | | | $1.04 \cdot 10^{-12}$ | |
| | Sonnad CFA orig | | | | | | | |
| [11] | Clamond 1it | (+) | 0 | 0.1 | 1000 | $10^{13}$ | n.a. | 3.7 |
| | Clamond orig | | | | | | $1 \cdot 10^{-15}$ (*) | |
| | Clamond 2it | | | | | | | |
| [12] | Biberg-14 | | 0 | 0.05 | 4000 | $10^8$ | $1.53 \cdot 10^{-3}$ | 3.7 |
| | Biberg-15 | | | | | | $6.1 \cdot 10^{-5}$ | |
| [16] | Vatankhah | (+) | 0 | 0.1 | 4000 | $10^8$ | n.a. | 3.71 |
| | Vatankhah orig | | | | | | $2.8 \cdot 10^{-5}$ | |
| [4] | Praks 18–28.1 | | 0 | 0.05 | 4000 | $10^8$ | $6.17 \cdot 10^{-4}$ | 3.7 |
| | Praks 18–28.2 | | | | | | | |
| | Praks 18–28.3 | | | | | | | |
| [26] | Praks 20a–2 | | 0 | 0.05 | 4000 | $10^8$ | $1.01 \cdot 10^{-3}$ | 3.71 |
| | Praks 20a–3 | | | | | | $3.66 \cdot 10^{-4}$ | |
| | Praks 20a–4 | | | | | | $8.08 \cdot 10^{-5}$ | |
| [15] | Praks 20b–29 | | 0 | 0.05 | 4000 | $10^8$ | $1.2 \cdot 10^{-5}$ | 3.71 |
| | Praks 20b–30 | | | | | | $2.4 \cdot 10^{-7}$ | |
| [27] | Lamri–14 | (+) | 0 | 0.1 | 4000 | $3 \cdot 10^8$ | $1.49 \cdot 10^{-3}$ | 3.7 |
| | Lamri–15 | | | | | | $4 \cdot 10^{-4}$ | |
| | Lamri–16 | | | | | | $2 \cdot 10^{-5}$ | |

# Appendix C: FORTRAN source code

For verification purpose the FORTRAN source code for the solution Clamond 1it opt is listed below.

```
FUNCTION CW_CLA1IT(Re,K)
! --------------------------------------------------------------------------------------------
! This FUNCTION computes the Weisbach friction factor for the Colebrook-White formula.
! Basis is the solution by source 2, with the optimized 1st iteration given in source 3).
! The maximal error is 2.79E-7, the accuracy is 6.55 MAD (minimal accurate digits).
!
! Input:    Re - Reynolds number, dimensionless
!           K - relative roughness, dimensionless
! Output:   Weisbach friction factor, dimensionless
!
! Sources:
! 1) Colebrook -1939
!    Turbulent flow in pipes, with particular reference to ... smooth and rough pipe laws
! 2) Clamond -2008
!    Efficient resolution of the Colebrook equation
! 3) Große-Dunker -2025
!    Improvement of Clamond’s solution of the Colebrook-White equation: highest accuracy ...
! --------------------------------------------------------------------------------------------
! Declaration of types
IMPLICIT NONE
! Input
REAL, intent(in):: Re,K
! Output
REAL            :: CW_CLA1IT
! Named constants
REAL, PARAMETER :: CK   =  3.7
REAL, PARAMETER :: CRe  =  2.51
REAL, PARAMETER :: one2 =  1.0/2.0
REAL, PARAMETER :: one3 =  1.0/3.0
REAL, PARAMETER :: A    = -2.0424324
REAL, PARAMETER :: C    = -6.0E-7
REAL, PARAMETER :: ApC  =  A + C               ! A+C=-2.0424330
REAL, PARAMETER :: P1   =  LOG(10.0)/(2.0*CK*CRe)
REAL, PARAMETER :: P2   =  LOG(Log(10.0)/(2.0*CRe))+A
REAL, PARAMETER :: P3   =  ((LOG(10.0)/2.0))**2.0
! Internal variables
REAL            :: X0, X1, E, R, Rp1
! --------------------------------------------------------------------------------------------
!  Begin of program
X0  = LOG(Re) + P2                              ! X0 - initial guess before first iteration
R   = K * Re * P1 + X0                          ! R = M + X0
Rp1 = R + 1.0                                   ! aux. variable Rp1 = R + 1
E   = (LOG(R) + ApC) / Rp1                      ! E for 1st iteration
X1  = X0 - (one2*E+Rp1)*E*R/(one3*E*E+E+Rp1)    ! X1 after first iteration
CW_CLA1IT = P3 / (X1 * X1)                      ! Weisbach friction factor
END FUNCTION CW_CLA1IT
```

## References

[1] C. F. Colebrook, "Turbulent flow in pipes, with particular reference to the transition region between the smooth and rough pipe laws", *Journal of the Institution of Civil Engineers*, vol. 11, no. 4, pp. 133–156, 1939, doi: 10.1680/ijoti.1939.13150.

[2] J. Weisbach, *Lehrbuch der Ingenieur- und Maschinen-Mechanik: Erster Theil: Theoretische Mechanik*. Braunschweig: Friedrich Vieweg und Sohn, 1845. [Online]. Available: https://www.google.de/books/edition/Lehrbuch_der_Ingenieur_und_Maschinen_Mec/gH1QAAAAcAAJ?hl=de&gbpv=1&dq=julius+weisbach+Lehrbuch+der+Ingenieur-+und+Maschinen-Mechanik+in+zwei+theilen+1845&printsec=frontcover&pli=1

[3] L. F. Moody, "Friction Factors for Pipe Flow", *Journal of Fluids Engineering*, vol. 66, no. 8, pp. 671–678, 1944, doi: 10.1115/1.4018140.

[4] P. Praks and D. Brkić, "Advanced Iterative procedures for solving the implicit Colebrook equation for fluid flow friction", *Advances in Civil Engineering*, vol. 2018, pp. 1–18, 2018, doi: 10.1155/2018/5451034.

[5] P. Praks and D. Brkić, "Choosing the optimal multi-point iterative method for the Colebrook flow friction equation", *Processes*, vol. 6, no. 8, p. 130, 2018, doi: 10.3390/pr6080130.

[6] C. T. Goudar and J. R. Sonnad, "Comparison of the iterative approximations of the Colebrook-White equation", *Hydrocarbon Processing: Fluid and Rotating Equipment, Special Report*, vol. 87, pp. 79–83, 2008.

[7] O. E. Turgut, M. Asker, and M. T. Coban, "A review of non iterative friction factor correlations for the calculation of pressure drop in pipes", *Bitlis Eren Univ J Sci Technol*, vol. 4, no. 1, 2014, doi: 10.17678/beujst.90203.

[8] L. Zeghadnia, J. L. Robert, and B. Achour, "Explicit solutions for turbulent flow friction factor: A review, assessment and approaches classification", *Ain Shams Engineering Journal*, vol. 10, no. 1, pp. 243–252, 2019, doi: 10.1016/j.asej.2018.10.007.

[9] L. E. Muzzo, D. Pinho, L. E. M. Lima, and L. F. Ribeiro, "Accuracy/speed analysis of pipe friction factor correlations" in *INCREaSE 2019: Proceedings of the 2nd International Congress on Engineering and Sustainability in the XXI Century*, J. Monteiro, Ed., Cham: Springer International Publishing AG, 2020, pp. 664–679.

[10] M. Niazkar and N. Talebbeydokhti, "Comparison of Explicit Relations for Calculating Colebrook Friction Factor in Pipe Network Analysis Using h-based Methods", *Iran J Sci Technol Trans Civ Eng*, vol. 44, no. 1, pp. 231–249, 2020, doi: 10.1007/s40996-019-00343-2.

[11] D. Clamond, *Efficient resolution of the Colebrook equation.* [Online]. Available: https://arxiv.org/abs/0810.5564v1

[12] D. Biberg, "Fast and accurate approximations for the Colebrook equation", *Journal of Fluids Engineering*, vol. 139, no. 3, 2017, doi: 10.1115/1.4034950.

[13] I. Santos-Ruiz, F.-R. López-Estrada, V. Puig, L. Torres, G. Valencia-Palomo, and S. Gómez-Peñate, "Optimal Estimation of the Roughness Coefficient and Friction Factor of a Pipeline", *Journal of Fluids Engineering*, vol. 143, no. 5, 2021, doi: 10.1115/1.4049674.

[14] P. Praks and D. Brkić, "Suitability for coding of the Colebrook's flow friction relation expressed by symbolic regression approximations of the Wright-ω function", *Rep. Mech. Eng.*, vol. 1, no. 1, pp. 174–179, 2020, doi: 10.31181/rme200101174p.

[15] P. Praks and D. Brkić, "Review of new flow friction equations: Constructing Colebrook's explicit correlations accurately", *RIMNI*, vol. 36, 2020, doi: 10.23967/j.rimni.2020.09.001.

[16] A. R. Vatankhah, "Approximate analytical solutions for the Colebrook equation", *J. Hydraul. Eng.*, vol. 144, no. 5, 2018, doi: 10.1061/(ASCE)HY.1943-7900.0001454.

[17] D. Brkić and P. Praks, "Accurate and efficient explicit approximations of the Colebrook flow friction equation based on the Wright ω-function", *Mathematics*, vol. 7, no. 1, p. 34, 2019, doi: 10.3390/math7010034.

[18] A. S. Householder, *The numerical treatment of a single nonlinear equation*. London: McGraw-Hill, 1970.

[19] N. Areerachakul, L. Girdwichai, and N. Areerakulkan, "Discussion of “Accurate explicit analytical solution for Colebrook-White equation” by Zahreddine Hafsi, Mechanics Research Communications 111 (2021) 103646)", *Mechanics Research Communications*, vol. 117, p. 103742, 2021, doi: 10.1016/j.mechrescom.2021.103742.

[20] P. K. Swamee and A. K. Jain, "Explicit equations for pipe-flow problems", *Journal of the Hydraulics Division*, vol. 102, pp. 657–664, 1976.

[21] N. H. Chen, "An explicit equation for friction factor in pipe", *Ind. Eng. Chem. Fund.*, vol. 18, no. 3, pp. 296–297, 1979, doi: 10.1021/i160071a019.

[22] D. J. Zigrang and N. D. Sylvester, "Explicit approximations to the solution of Colebrook's friction factor equation", *AIChE J.*, vol. 28, no. 3, pp. 514–515, 1982, doi: 10.1002/aic.690280323.

[23] S. E. Haaland, "Simple and explicit formulas for the friction factor in turbulent pipe flow", *Journal of Fluids Engineering*, vol. 105, no. 1, pp. 89–90, 1983, doi: 10.1115/1.3240948.

[24] T. K. Serghides, "Estimate friction factor accurately", *Chemical Engineering Journal*, vol. 91, pp. 63–64, 1984.

[25] J. R. Sonnad and C. T. Goudar, "Explicit reformulation of the Colebrook−White equation for turbulent flow friction factor calculation", *Ind. Eng. Chem. Res.*, vol. 46, no. 8, pp. 2593–2600, 2007, doi: 10.1021/ie0340241.

[26] P. Praks and D. Brkić, "Accurate and efficient explicit approximations of the Colebrook flow friction Equation based on the Wright ω-function: Reply to the discussion by Majid Niazkar", *Mathematics*, vol. 8, no. 5, p. 796, 2020, doi: 10.3390/math8050796.

[27] A. A. Lamri and S. M. Easa, "Computationally efficient and accurate solution for Colebrook equation based on Lagrange theorem", *Journal of Fluids Engineering*, vol. 144, no. 1, 2022, doi: 10.1115/1.4051731.